\documentstyle[prl,aps,epsf]{revtex}

\begin{document}

\twocolumn[
\hsize\textwidth\columnwidth\hsize\csname@twocolumnfalse\endcsname 

\title{Active suppression of dephasing in Josephson-junction qubits} 

\author{D.V. Averin$^{(1)}$ and Rosario Fazio$^{(2)}$}

\address{ $^{(1)}$Department of Physics and Astronomy, SUNY 
Stony Brook, Stony Brook, NY 11794-3800  \\
$^{(2)}$NEST-INFM $\&$ Scuola Normale Superiore, 56126	Pisa, 
Italy} 

\date{\today} 

\maketitle  

\begin{abstract} 

Simple majority code correcting $k$ dephasing errors by encoding 
a qubit of information into $2k+1$ physical qubits is studied 
quantitatively. We derive an equation for quasicontinuous evolution 
of the density matrix of encoded quantum information under the 
error correction procedure in the presence of dephasing noise that 
in general can be correlated at different qubits. Specific design 
of the Josephson-junction circuit implementing this scheme is 
suggested.

\end{abstract}

\pacs{PACS numbers: 03.67.Pp,03.65.Yz,73.23.Hk}

]

Josephson qubits are among the most promising devices to implement 
solid state quantum computation~\cite{b1,b2}. Quantum manipulations 
of individual~\cite{b3,b4,b5,b6,b7,b8,b9} and coupled~\cite{b10} 
qubits has been demonstrated experimentally. At present, probably 
the main obstacle to the development of a larger-scale solid state 
quantum logic circuits is presented by decoherence. It is therefore 
important to develop strategies to minimize the effects of 
decoherence on the dynamics of the qubit systems.

Known approaches to reduction of decoherence include both 
error-correction and error-avoiding schemes that either employ 
symmetries of the qubit-environment interaction to create areas of 
the Hilbert space not affected by decoherence~\cite{b11,b12} or use 
rapid random dynamic perturbations of the system to average out the 
effects of external noise~\cite{b13,b14}. The error-avoiding 
approaches appear  to be less suitable for the solid-state qubits. 
Indeed, noise in solid-state systems typically does not have any 
particular symmetry and its correlation time is short, so that the 
application of the control pulses within this time-scale, as 
required by the dynamic averaging schemes, is problematic. This 
leaves error-correction as the main strategy for suppression of 
decoherence in solid-state qubits. In this work, we suggest 
an implementation of one of the basic error-correction 
algorithms for the suppression of dephasing errors (which can be 
expected to be the dominant type of errors in solid-state circuits 
- see, e.g., Ref.~\cite{b9}), and develop its quantitative 
description. Our scheme employs the Josephson-junction qubits that 
combine charge and flux dynamics~\cite{b15,b17,b6}, and requires 
only a small number of qubit transformations to operate. 

From the perspective of the general theory of error-correction, 
an interesting feature of the scheme considered in this work is 
the possibility of developing its detailed quantitative 
description within the realistic model of the qubit-environment 
interaction and analyzing, for instance, the effect of the 
correlations in the noise acting on different qubits. While 
discussions of the error-correction rely typically in an 
essential way on independent noise models, environments of the 
solid-state qubits can be to a large degree correlated because 
of the finite distance between the qubits in a circuit. A clear 
illustration of this is provided by the background charge 
fluctuations that are the main source of dephasing in charge 
qubits~\cite{b3,b6,b10,b16}. Long-range nature of the Coulomb 
interaction creates noise correlations by coupling the qubits 
to the same charge fluctuators. 

We consider specifically the problem of ``quantum memory'', when 
the task is to preserve the stationary state of the qubit in the 
presence of dephasing noise. The qubit Hamiltonian contains then 
only the coupling to the environment. Under the assumption 
that the environment has many degrees of freedom each of which is 
only weakly coupled to qubits, it can be modeled as an ensemble 
of harmonic oscillators~\cite{b18,b19,b20,b21} (see however 
\cite{b16}), so that the Hamiltonian of the qubit register 
is: 
\begin{equation}
H=\sum_{j} \sigma_z^{(j)}\xi_j \, , 
\label{1} 
\end{equation} 
where $\xi_j= \sum_{m,k} [\lambda_{m,j}(\omega) a_{m,\omega} + 
\mbox{h.c.}]$. 
Here we assumed several independent ensembles of environmental 
oscillators (numbered by $m$), as needed to model different 
profiles of spatial correlations of random forces $\xi_j$. The 
index $j=1,2... $ in~(\ref{1}) numbers the qubits, 
and coefficients $\lambda_{m,j}(\omega)$ are coupling constants 
of the qubit $j$ to the oscillators of reservoir $m$ in the mode
$\omega$ and creation/annihilation operators $a_{m,\omega}, 
a^{\dagger}_{m,\omega}$. Time evolution of the 
``qubits+environment'' system is described conveniently in the 
interaction representation with respect to the interaction 
Hamiltonian of Eq.\ (\ref{1}). The evolution operator $U(t)$ 
can then be calculated explicitly by separating the two 
non-commuting parts, $a_{m,\omega}$, and $a^{\dagger}_{m,\omega}$, 
of the qubit-oscillator coupling, and using the fact that their 
commutator is a $c$-number: 
\begin{equation} 
U(t) = \exp \{ -i \sum_{j} \varphi_j(t) \sigma_z^{(j)} \} U_r(t) 
\, , 
\label{2} \end{equation}

\vspace{-3ex}

\[ U_r(t) = \exp \{ i \sum_{m,\omega} \frac{\omega t-\sin 
\omega t}{\omega^2} | \sum_{j} \lambda_{m,j}\sigma_z^{(j)} |^2 
\} \, . \]
The first term in $U(t)$ represents fluctuating phases 
$\varphi_j(t)$ of the qubit basis states induced by the 
environmental forces $\xi_j(t)$: $\varphi_j(t) =\int_0^{t} 
\xi_j(t') dt'$. The second term, $U_r(t)$, results from the 
renormalization of the qubit parameters by the qubit-environment 
interaction. To see this more explicitly, we note that 
the sum over frequencies $\omega$ in this exponent has a natural 
cut-off at large frequencies $\omega \simeq \tau_c^{-1}$, where 
$\tau_c$ is the time scale at which environment forces acting 
on different qubits are correlated. For weak decoherence we are 
interested in the time scales much larger that $\tau_c$. In 
this regime, the phase represented by $U_r(t)$ is dominated by 
the term that grows linearly with $t$, and can be viewed 
as arising from the renormalization of the qubit energy. 
Equation (\ref{2}) shows that such a renormalization includes 
then the shift of the total energy of the register and creation 
of the qubit-qubit interaction. The total energy shift is 
irrelevant as long as we consider an individual register. 
Neglecting it, we see that $U_r(t)$ results from the Hamiltonian 
evolution with the Hamiltonian 
\begin{equation}
H_r =  -\sum_{j,j'}V_{jj'}\sigma_z^{(j)}\sigma_z^{(j')} \, , 
\label{3} \end{equation} 
and $V_{jj'} = 2\mbox{Re} \sum_{m,\omega} (\lambda_{m,j}(\omega) 
\lambda^*_{m,j'}(\omega)/ \omega)$, if the sum over frequencies 
$\omega$ in this expression is converging at low frequencies. The 
qubit-qubit interaction strength $V_{jj'}$ is non-vanishing only 
if the same reservoir $m$ couples to more than one qubit, so that 
the reservoir forces $\xi_j$ at different qubits are correlated. 

The time evolution with the Hamiltonian $H_r$, and more 
generally, the evolution operator $U_r$ in Eq.\ (\ref{2}) 
represent deterministic part of the qubit evolution induced by 
the qubit-reservoir interaction. As a result, it can in principle 
be compensated for by adjusting the regular (non-dissipative) 
part of the Hamiltonian of the qubit register. This procedure, 
however, is impractical even in the case of constant $H_r$, since 
the interaction constants $V_{jj'}$ are apriori unknown and 
incommesurate quantities. This complexity means that a more 
appropriate approach is to treat the time evolution represented 
by $U_r$ as dephasing despite its deterministic character. 

The time evolution of the density matrix $\rho (t)$ of the 
qubit register is obtained from Eq.\ (\ref{2}) through the 
relation $\rho (t)=\mbox{Tr}_{env} \{ U^{\dagger} (t) 
\sigma (0) U(t) \}$, where $\sigma$ is the total density 
matrix of the ``qubits+environment'' system. The environment 
will dephase the qubits if they are prepared initially in the 
state $\rho(0)$ that is uncorrelated with the state of the 
environment, $\sigma (0) = \rho_{env} \rho(0)$. Assuming that the 
environment is in thermal equilibrium at temperature $\Theta$, 
and no error correction procedure is applied, we get using the 
standard property of the Gaussian noise: 

\[ \rho(t) = \exp \{ -\frac{1}{2} \sum_{j,j'} \langle 
\varphi_j(t) \varphi_{j'}(t)\rangle  (\sigma_z^{(j)}- \bar{ 
\sigma}_z^{(j)}) (\sigma_z^{(j')}- \bar{\sigma}_z^{(j')})\} \]

\vspace{-3ex}

\begin{equation}
\cdot U_r^{\dagger} (t) \rho(0) U_r(t) \, . 
\label{4} 
\end{equation}
Here we introduced the convention that the bar over $\sigma_z$ 
operators means that they act on $\rho$ from the right. 
Qualitatively, Eq.\ (\ref{4}) shows that the matrix elements of 
$\rho$ that are further away from the diagonal in the $\sigma_z$ 
basis decay faster. The diagonal elements (on which $\sigma_z- 
\bar{\sigma}_z=0$) remain constant. In the case of one physical 
qubit, Eq.\ (\ref{4}) gives $\rho (t) = e^{-\langle \varphi^2(t) 
\rangle (1-\sigma_z \bar{\sigma}_z)} \rho(0)$, i.e., the 
off-diagonal elements of $\rho$ are suppressed with time as 
$e^{-2 \langle \varphi^2(t)\rangle } \equiv e^{-P(t)}$.  
If the environment density of states is Ohmic, i.e., $\sum_{m, 
\omega} |\lambda_m (\omega)|^2 ...  = g \int_0^\infty  d\omega 
\omega e^{-\omega \tau_0} ... \, $, direct evaluation for 
$\Theta \ll 1/\tau_0$ gives: $P(t)  = 2g \ln [\sinh (\pi t 
\Theta)/(\pi \tau_0 \Theta) ]$. At large $t$, when the random 
force $\xi$ appears $\delta$-correlated, $P(t)$ reduces to 
$P(t)= \Gamma t$, where $\Gamma= 2\pi g \Theta$ is the 
dephasing rate. 

One can reduce the effective dephasing rate by the encoding 
that corrects the phase errors~\cite{b22,b23}. Generalized to 
$k$ errors, this encoding is: 
\begin{equation}
\alpha |0\rangle +\beta |1\rangle \rightarrow \alpha | 
\underbrace{++...+}_{2k+1} \rangle + \beta |\underbrace{--...- 
}_{2k+1} \rangle \, . 
\label{6} \end{equation}
In Equation (\ref{6}), a bit of quantum information is encoded 
in the state of the $2k+1$ physical qubits, and the $|\pm\rangle$ 
states of each of these qubits are obtained through the Hadamard 
transform $\hat{H}$ (the $\pi/2$-rotation around $y$ axis) from 
the $|0,1\rangle$ states. All of the $\sigma_z$ operators in the 
dephasing-induced time evolution (\ref{2}) are changed 
by $\hat{H}$: $\hat{H}\sigma_z \hat{H} =\sigma_x$, so that for 
the states on the right-hand-side of Eq.\ (\ref{6}) the dephasing 
looks like transitions between the $|\pm\rangle$ states of each 
qubit, and can be directly detected by measurements in this basis 
and corrected by applying simple pulses returning the qubit into 
the initial state. The error-detecting measurements, however, 
should not destroy the quantum information encoded in the state 
(\ref{6}), i.e., they should not distinguish the $\alpha$ and 
$\beta$ parts of this state. This condition is not satisfied by 
measurements on individual qubits but can be satisfied by the 
measurements on pairs of the nearest-neighbor qubits comparing 
their states. Despite the apparent complexity of this scheme, it 
has quite natural implementation in the Josephson-junction qubits 
- see Fig. 1. 

To describe this process quantitatively we assume that its 
measurement/correction part can be done on the time scale that 
is much shorter than the one set by the characteristic dephasing 
rate $\Gamma$. Different terms in the environment-induced 
evolution of the encoded state, Eq.\ (\ref{6}), during the time 
interval $T$ between the successive application of the 
``measurement+correction'' operations can be conveniently 
classified by the number of qubits flipped during this time 
interval. In the relevant regime of sufficiently short $T$: 
$P(T) \ll 1$, the probability amplitude of these terms decreases 
rapidly when this number increases. If we keep only the terms 
that flip up to $k$ qubits, we see directly from Eq.\ (\ref{2}) 
that the time evolution at this level of accuracy (denoted by 
$U_k(T)$) preserves the superposition of the $\alpha$ and 
$\beta$ parts of the encoded state: 
\begin{equation}
U_k(T)[ \alpha | \oplus \rangle +\beta  |\ominus \rangle 
]= \sum_{q} [\alpha |\psi_q \rangle +\beta \hat{R} |\psi_q 
\rangle ] u_q \, . 
\label{7} \end{equation}
Here index $q$ runs over $2^{2k}$ different register states 
obtained from the state $|\oplus\rangle \equiv |+...+\rangle $ 
by flipping up to $k$ qubits, $u_q$ are the probability 
amplitudes of these states, $\hat{R}|\psi_q \rangle$ denotes 
the state $|\psi_q \rangle$ with all $2k+1$ qubits inverted, 
and $|\ominus \rangle \equiv |-...-\rangle $. 

The measurements that compare the qubit states in all pairs 
of the nearest-neighbor qubits do not distinguish states 
$|\psi_q \rangle$ and $\hat{R}|\psi_q \rangle$, and therefore 
also preserve the superposition of the $\alpha$ and $\beta$ 
terms in Eq.\ (\ref{7}). The $2^{2k}$ different outcomes 
(``equal'' 
or ``different'') of the $2k$ such measurements distinguish 
all terms with different $q$ in Eq.\ (\ref{7}) and enable one 
to decide what qubits were flipped during the time interval 
$T$. Application of the correcting pulses should then bring 
the state of the qubit register back to its initial form 
(\ref{6}) so that the encoded quantum state does not change 
in this approximation. The residual evolution of the encoded 
state is associated with the possibility that environment flips 
more that $k$ different qubits; for $P(T)\ll 1$ -- precisely 
$k+1$ qubits. Following the same steps as above, we see that 
when $k+1$ qubits are flipped, the measurement/correction cycle 
interchanges the $\alpha$ and $\beta$ weights in the encoded 
state (\ref{6}). Since the probability $p$ of this mistake is 
small, $p\ll 1$, the encoded state changes substantially only 
on the time scale larger than the period $T$ of one 
error-correction cycle, and its evolution on this scale can be 
conveniently described by the continuous equation for the 
density matrix $\rho^{(c)}$ in the basis of $|\oplus \rangle$ 
and $|\ominus \rangle$ states. The interchange of the $\alpha$ 
and $\beta$ terms in (\ref{6}) leads to the following evolution 
of $\rho^{(c)}$:  
\begin{equation} 
\dot{\rho}^{(c)}_{++}=\frac{\gamma_k}{2} (\rho^{(c)}_{--}
			-\rho^{(c)}_{++}) \, ,\;\;\;\; 
\dot{\rho}^{(c)}_{+-} =\frac{\gamma_k}{2} 
			(\rho^{(c)}_{-+} -\rho^{(c)}_{+-}) \, .  
\label{8} \end{equation} 
Here $\gamma_k \equiv 2p/T$ is the effective dephasing rate of 
the encoded quantum information, and the superscript $(c)$ 
indicates 
that $\rho^{(c)}$ is the reduced density matrix in the presence 
of error correction. Thus, our error-correcting procedure replaces 
the dephasing in the individual physical qubit with the dephasing 
of encoded quantum information at a smaller rate. Indeed, if one 
writes Eqs.\ (\ref{8}) in the rotated basis $|\oplus \rangle \pm 
|\ominus \rangle$, they explicitly acquire the form characteristic 
for pure dephasing: constant diagonal elements of the density matrix 
and decay of the off-diagonal elements with the rate $\gamma_k$. 
The dephasing rate $\gamma_k$ can be calculated from the evolution 
operator (\ref{2}). Now we discuss several important limits. 

For  $k=1$, when the relevant errors flip 2 out of 3 qubits, we 
get: 
\[ \gamma_1 = \frac{2}{T} \sum_{j>j'} (T^2V_{jj'}^2+ 2 \langle 
\varphi_j(T) \varphi_{j'}(T)\rangle^2 + \langle \varphi_j^2(T) 
\rangle  \langle \varphi_{j'}^2(T)\rangle ) \, . \]   
The first two terms in this expression represent contribution 
to dephasing from noise correlations at different qubits, 
while the last term exists also for uncorrelated noise. If the 
noise is $\delta$-correlated in time, $\gamma_1$ reduces to 
$\gamma_1 = T\sum_{j>j'} (2V_{jj'}^2 +\Gamma_{jj'}^2 +\Gamma_j 
\Gamma_{j'}/2)$, where $\Gamma_j$ is the dephasing rate 
in the $j$th qubit, and $\Gamma_{jj'}$ is introduced through 
$2\langle \varphi_j(t) \varphi_{j'}(t)\rangle = \Gamma_{jj'}t$. 

For $k=2$, the dephasing rate of the encoded state is: 
\begin{equation} 
\gamma_2 = \frac{2}{T} \sum_{j>j'>j''} \langle \varphi_j^2 
\varphi_{j'}^2 \varphi_{j''}^2\rangle \, .   
\label{13} 
\end{equation} 
Since the effective coupling induced by the environment -- see 
Eq.\ (\ref{3}), flips the qubits only in pairs, it does not 
contribute to $\gamma_2$. If the dephasing noise is 
$\delta$-correlated in time, its space correlations are 
non-vanishing only for the nearest-neighbor qubits, and the 
corresponding dephasing rates are the same for all qubits, 
Eq.\ (\ref{13}) gives: 
$\gamma_2 = 5 \Gamma T^2 (\bar{\Gamma}^2 + \Gamma^2/2)$, where 
$\bar{\Gamma} \equiv \Gamma_{jj+1}$. 

If the dephasing forces at different qubits are uncorrelated, 
the encoded dephasing rate can be easily calculated for 
arbitrary $k$: 
\begin{equation}
\gamma_k = \frac{1}{2^kT} \sum_{j_1>j_2> ...…>j_{k+1}} 
P_{j_1}(T) P_{j_2}(T) ... P_{j_{k+1}}(T) \, ,   
\label{9} 
\end{equation} 
and one sees that $\gamma_k$ decreases exponentially with the 
``degree of encoding'' $k$. When the probabilities $P(T)$ of 
dephasing errors in individual qubits can be expressed through 
the dephasing rate $\Gamma$, Eq.\ (\ref{9}) reduces to 
$\gamma_k = \Gamma (\Gamma T)^k (2k+1)! /(2^k k!(k+1)!)$, if 
$\Gamma$ is the same for all qubits.
 
Exponential suppression of $\gamma_k$ with $k$ is limited 
in the scheme considered above by possible imperfections of 
the measurement/correction operations. The most important 
is direct dephasing of the encoded state by measurements, 
which, in contrast to correction steps, need to be performed 
each period $T$. For example, one of the specific non-idealities 
of measurement detectors that leads to direct dephasing of the 
encoded state is residual linear response coefficient of the 
quadratic detectors needed to perform pair-wise comparison of 
the qubit states -- see Eq.\ (\ref{15}) below. Linear terms 
couple the detector directly to the $|\pm \rangle$ states of 
individual qubits and introduce finite phase shifts between them. 
Since the number of required measurements is proportional to $k$, 
the rate of introduced dephasing should also be proportional to 
$k$, and can be denoted as $\bar{\gamma} k$. The effect of this 
dephasing on the evolution of the encoded quantum information is 
described then by adding the usual dephasing term to the equation 
for the off-diagonal element of the density matrix $\rho^{(c)}$ 
(\ref{8}): 
\begin{equation}
\dot{\rho}^{(c)}_{+-} = \frac{\gamma_k}{2} (\rho^{(c)}_{-+}- 
\rho^{(c)}_{+-}) -\bar{\gamma} k \rho^{(c)}_{+-}\, .  
\label{14} \end{equation} 
Qualitatively, the two types of dephasing processes in Eq.\ 
(\ref{14}) have similar effect of suppressing the fidelity of 
the encoded state, but depend differently on $k$. The optimum 
degree of encoding is estimated crudely by minimizing the total 
dephasing rate: $k_{opt} \sim \ln (\bar{\gamma}/\Gamma)/\ln 
(T\Gamma)$. One obvious result of this optimization is that for 
the considered scheme of the dephasing suppression to make 
sense, the dephasing introduced by imperfections of the 
correcting procedure should be much weaker than the original 
qubit dephasing $\Gamma$. 

\begin{figure}[htb]
\setlength{\unitlength}{1.0in}
\begin{picture}(3.2,2.15) 
\put(.25,.1){\epsfxsize=2.7in\epsfbox{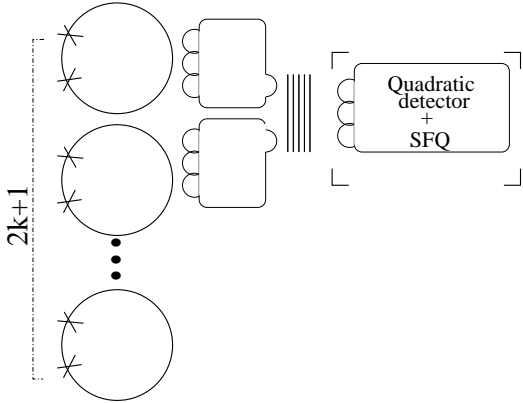}}
\end{picture}
\caption{Schematics of the Josephson-junction circuits 
implementing suppression of dephasing. Crosses denote 
tunnel junctions, the electrodes between them act as 
charge qubits. Monitored currents in the nearest-neighbor 
loops enclosing qubits allow to detect dephasing errors.} 
\end{figure} 

This condition can be satisfied in Josephson-junction 
qubits, where the dynamics of magnetic flux characterized by 
longer coherence times (at least tens of nanoseconds -- see, 
e.g., \cite{b9}) can be used to suppress dephasing in 
charge-based qubits. The charge qubits have quite short 
decoherence times, $\sim 1$ ns~\cite{b3,b10}, limited by the 
background charge fluctuations, but offer some advantages, 
e.g., demonstrated simplicity of the qubit-qubit 
coupling~\cite{b10}. Therefore it would be of interest to use 
the approach discussed in this work to suppress dephasing of 
charge degrees of freedom with the help of controlled flux 
dynamics. A sketch of the possible set-up achieving this is 
shown in Fig.\ 1. Its main elements are the charge qubits, 
formed by two small tunnel junctions in series, enclosed 
in small superconducting loops threaded by magnetic flux 
$\Phi$ equal to half of the magnetic flux quantum $\Phi_0$. 
It can be shown~\cite{b17} that the current in each loop 
represents the  $\sigma_x$ component of the qubit 
dynamics, and its monitoring measures therefore the qubit 
in the $\sigma_x$ basis  as needed for detection of the 
dephasing errors. Comparison of the states of the 
nearest-neighbor qubits can be achieved by measuring not 
directly the currents in the loops, but the square of the 
difference (or of the sum) between the currents. Such a 
quadratic detection measures the product operator 
$\sigma_x^{(j)}\sigma_x^{(j+1)}$: 
\begin{equation} 
(\sigma_x^{(j)} \pm \sigma_x^{(j+1)})^2= 2(1 \pm 
\sigma_x^{(j)}\sigma_x^{(j+1)}) \, , 
\label{15} \end{equation}
and provides information on whether the states of the two 
qubits are the same or not without measuring them. Quadratic 
detection can be realized by the usual magnetometers but 
operated at a point where the linear response coefficient 
vanishes. These measurements, subsequent classical 
calculations, and application of correction pulses, can be done 
with sufficient frequency by existing ``SFQ'' superconductor 
electronics~\cite{b24} compatible with the qubits. 

In summary, we suggested a simple scheme of performing basic 
error-correction in Josephson-junction qubits. The scheme 
suppresses dephasing errors and can be analyzed quantitatively 
within the realistic model of environment, including the 
possibility of noise correlations at different qubits. If the 
errors introduced by the correction procedure are negligible, 
the residual dephasing rate for the encoded quantum 
information decreases exponentially with the degree of 
encoding.  

\vspace*{.4ex}

This work was supported in part by the NSF under grant \# 
0121428, and by the NSA and ARDA under the ARO contract 
(D.V.A.), and by EC-grant IST-FET-SQUBIT (R.F.)

\end{document}